\documentclass[doublespacing]{elsart}
\usepackage{latexsym,amsmath,tabularx,amssymb,bm}
\usepackage{epsfig}
\usepackage{citesort}

\usepackage{graphicx}
\usepackage{dcolumn}
\usepackage{bm}

\newcommand \beq{\begin{eqnarray}}
\newcommand \eeq{\end{eqnarray}}
\newcommand \bea{\begin{eqnarray}}
\newcommand \eea{\end{eqnarray}}

\newcommand \bk{{\mathbf k}}

\usepackage{amsmath}
\usepackage{amssymb}

\def\simle{\mathrel{
          \rlap{\raise 0.511ex \hbox{$<$}}{\lower 0.511ex \hbox{$\sim$}}}}

\begin{document}
\begin{flushright}
~\vspace{-1.25cm}\\
{\small\sf ECT*-05-22, SPhT-T06-010, UT-Komaba/06-3}
\end{flushright}
\vspace{0.8cm}
\begin{frontmatter}
\parbox[]{16.6cm}{ \begin{center}

\title{Landau-Pomeranchuk-Migdal effect in a quark-gluon plasma
and the Boltzmann equation}

\author[a]{Gordon Baym},
\author[ect]{J.-P. Blaizot\thanksref{cnrs}},
\author[c]{F. Gelis}, and \author[d]{T. Matsui}
\address[a]{Department of Physics,  University of Illinois, 1110 W. Green
St., Urbana, IL, 61801 USA}
\address[ect]{$ECT*$, Villa Tambosi, Strada delle Tabarelle 286,
I-38050 Villazzano(TN), Italy}
\address[c]{SPhT, CEA-Saclay, 91191 Gif-sur-Yvette cedex, France}
\address[d]{Institute of Physics, University of Tokyo, Komaba,
Tokyo, Japan}
\thanks[cnrs]{Membre du Centre National de la Recherche Scientifique
(CNRS), France.}

\vspace{0.8cm}

\begin{abstract}

     We show how the Landau-Pomeranchuk-Migdal effect on photon production
rates in a quark-gluon plasma can be derived via the usual Boltzmann equation.
To do this, we first derive the electromagnetic polarization tensor using
linear response theory, and then formulate the Boltzmann equation including
the collisions mediated by soft gluon exchanges.  We then identify the
resulting expression for the production rate with that obtained by the
field-theoretic formalism recently proposed by Arnold, Moore and Yaffe.  To
illustrate the LPM effect we solve the Boltzmann equation in the diffusion
approximation.

\end{abstract}
\end{center}}
\end{frontmatter}

\newpage

    The Landau-Pomeranchuk-Migdal (LPM) effect \cite{LP53,Mig56} plays an
important role in the diagnostic tools of quark-gluon plasmas formed in
ultrarelativistic nucleus-nucleus collisions, both in the energy loss
\cite{BDPS95}, and the emission of photons and dileptons from the plasma, our
focus here (for a recent review see \cite{GELIS-Quark Matter,Blaizot:2005mj}).
The LPM effect takes into account multiple scatterings of the emitters and
subsequent interference of the emitted radiation leading to a suppression of
the bremsstrahlung rate from that obtained by Bethe and Heitler \cite{BH34}.

    In recent discussions \cite{ArnolMY123}, the LPM effect has been
calculated by explicitly summing the infinite series of Feynman diagrams that
correspond to multiple scatterings.  While providing a modern
field-theoretical derivation of Migdal's results, such an approach is
complicated by the need first to identify the relevant series of diagrams, and
then to approximate these diagrams carefully in order to obtain useful
expressions.  As in the Landau theory of Fermi liquids \cite{bp}, multiple
scattering processes involving a sequence of singular denominators are most
effectively dealt with in the framework of the Boltzmann equation:  not only
does the Boltzmann equation capture the relevant diagrams, it also has the
necessary kinematical approximations for small-angle scatterings built in via
the gradient expansion of the collision term.  However, neither the field
theoretic treatment of \cite{ArnolMY123}, nor in fact Migdal's original
derivation, make manifest the fact that the effect of multiple collisions is
entirely captured in the relevant kinematical regime by the usual linearized
Boltzmann equation.\footnote{The first and third references in \cite
{ArnolMY123} do note an integral equation analogous in structure to a
Boltzmann equation, but this analogy is not explored further.  The Boltzmann
equation derived here should not be confused with that discussed in
\cite{Arnold:2001ms}:  in the latter paper, the LPM effect enters as a
correction to the $1\to 2$ and $2\to 1$ collision terms.  Migdal's original
derivation describes the propagation of a charged particle interacting with
fixed scattering centers at random locations.  In a sense, the derivation
presented in this paper extends Migdal's work \cite{Mig55} to the case where
scatterings are due to two-body collisions.  The present generalization
focusses on soft photons, while Migdal's approach is valid for arbitrary
photon energies.} Establishing this simple connection is the main purpose of
the present Letter; we derive the LPM effect based on a (linearized) Boltzmann
equation.  While the rates obtained by solving the Boltzmann equation are not
different from those obtained by the formalism of \cite{ArnolMY123}, the use
of the Boltzmann equation significantly simplifies the derivation of the LPM
effect and provides a new, more intuitive, perspective on the problem.
Moreover, it is also the basis of new tools for addressing this issue in a
non-equilibrated plasma.

    To leading order in the electromagnetic fine structure constant, $\alpha$,
the photon production rate is \cite{Weldo3,GaleK1}:
\begin{eqnarray} {{\omega\frac{dN_\gamma}{d^4 x\, d^3\bk}}} =-\frac{
  g^{\mu\nu}}{2(2\pi)^3} \Pi^<_{\mu\nu}(\omega,\bk)\;.
  \label{eq:rate-1}
  \end{eqnarray}
Here $\omega=|\bk|$, and $\Pi^<_{\mu\nu}(\omega,\bk)$ is the Fourier
transform of the finite temperature current-current correlation function
($K\equiv(\omega,\bk)$):
\begin{eqnarray}
\label{currentcurrent}
\Pi^<_{\mu\nu}(\omega,\bk)\equiv\int{\rm d}^4
X\;{\rm e}^{iK\cdot X}\;\langle j_\mu(0)j_\nu(X)\rangle\; ,
\end{eqnarray}
with $j_\mu(X)\equiv e \bar\psi(X)\gamma_\mu \psi(X)$ the electromagnetic
current and $X\equiv (t,{\bf x})$ denotes the space-time coordinates.  We use
a metric with $g^{00}=1$.  To derive (\ref{eq:rate-1}) we use the
transversality of $\Pi_{\mu\nu}^<(K)$ :  $k^\mu \Pi_{\mu\nu}^<(K)=0$.  A
similar formula exists for lepton pairs for which $K^2\equiv
\omega^2-\bk^2>0$.  For real photons, one can replace
$g^{\mu\nu}\Pi_{\mu\nu}^<$ by $g^{ij}\Pi_{ij}^<$, where $i,j=1,2$ are the two
directions transverse to the photon momentum; indeed the transversality of
$\Pi_{\mu\nu}^<$ ensures that non-transverse polarizations do not contribute.
The brackets in Eq.~(\ref{currentcurrent}) denote a thermal average.  The
correlation function $\Pi^<_{\mu\nu}(\omega,\bk)$ is related to the retarded
electromagnetic polarization tensor through (see, e.g., \cite{KB61,LeBellac}),
\bea
       \Pi^<_{\mu\nu}(\omega,\bk)= -
       \frac{2}{e^{\omega/T}-1} \;{ {\rm Im}\,\Pi^{\rm
        ret}_{\mu\nu}(\omega,\bk)}\; .
\eea
For $\omega \ll T$ the photon production rate per unit volume and
frequency is thus
\begin{eqnarray}
    \frac{dN_\gamma}{d^4 x d\omega} = -\frac{T}{2\pi^2}\sum_{i=1}^{2}
    \;{{\rm Im}\,\Pi^{\rm ret}_{ii}(\omega,\bk)}\;.
   \label{rate-2}
\end{eqnarray}
The main task in estimating the photon production rate is therefore to
calculate $\Pi_{ij}^{\rm ret}(\omega,\bk)$.  Since we can write the
self-energy in terms of the response of the electromagnetic current to an
external electromagnetic vector potential,
\bea
       \Pi_{\rm  ret}^{ij} =\frac{\delta \langle j^i\rangle}{\delta A_j},
       \label{deltajA}
\eea
the problem reduces to calculating $\langle j^i\rangle$ in the presence
of an external $A_j$.

\begin{figure}
\begin{center}
\resizebox*{!}{3.0cm}{\includegraphics{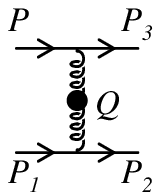}}\hglue 30mm
\resizebox*{!}{3.5cm}{\includegraphics{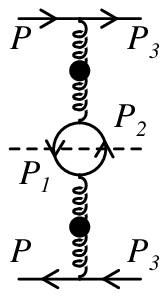}}
\end{center}
\caption{\label{Born} Left:  quark-quark elastic scattering in the Born
approximation via gluon exchange.  The straight lines denote quarks, and the
curly lines gluons.  Dynamical screening corrections, denoted by the thick
dot, can be included via the hard thermal loop expansion.  Right:
the cut on the internal quark loop bubble that is used in connection with
Eq.~(\ref{gammaC2}) below.Note that the scattering partner, as well the
excitation running in the cut loop, can also be a gluon.}
\end{figure}

\begin{figure}[htb]
\begin{center}
\resizebox*{8.4cm}{!}{\includegraphics{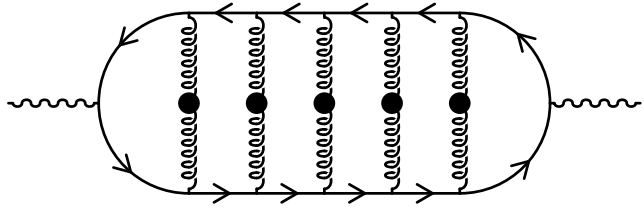}}
\end{center}
\caption{\label{fig:equations}   Resummation of ladder diagrams in
the photon polarization tensor taken into account by the Boltzmann equation.
The thick dot denotes dynamical screening corrections.
}
\end{figure}

    As we show below, the physics of the LPM effect is included in a
calculation of the response of the current to an external field via the
Boltzmann equation, even with the simple collision term describing scatterings
at the Born approximation level (see Fig.~\ref{Born}).  The solution of the
Boltzmann equation takes into account repeated scatterings -- processes needed
to include the physics of the LPM effect -- as illustrated in
Fig.~\ref{fig:equations}.

     The leading order contributions to the photon rate, of order
$\alpha_s$ ($\equiv g^2/4\pi$, with $g$ the strong coupling
constant), correspond to real gluon-photon Compton scattering ($qg\to
q\gamma$ or $\bar{q}g\to \bar{q}\gamma$) and quark-antiquark
annihilation ($q\bar{q}\to g\gamma$).  These processes can be
calculated by including hard thermal loop (HTL)
\cite{BraatP1,FrenkT1} corrections in the propagators
\cite{KapusLS1,BaierNNR1}, and do not require further resummation.
In particular they are not affected by the LPM effect.  Since we
focus here on the LPM effect we omit out these processes in the
following discussion.

\begin{figure}[htb]
\centering \resizebox*{!}{2.5cm}{\includegraphics{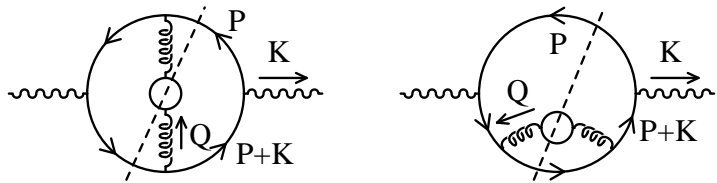}}
\caption{\label{fig:processes-2} Higher order processes that are promoted
to ${\mathcal O}(\alpha_s)$ by collinear singularities.}
\end{figure}

     The processes shown in Fig.~\ref{fig:processes-2}, formally of next order
in $\alpha_s$, are collisions involving space-like gluons.  The one-loop
correction in the gluon propagator shown in these three-loop processes is only
the first correction; the full correction should be carried out in practice by
an HTL resummation.  Naive power counting suggests that these two diagrams
contribute in ${\mathcal O}(\alpha_s^2)$; however, ``collinear enhancement''
turns the contribution of these diagrams into a contribution of order $
\alpha_s$.

     We illustrate the origin of this enhancement by first studying the extent
to which the quark of momentum $P+K$, between the quark-gluon vertex
and the photon emission vertex in Fig.~\ref{fig:processes-2}, is
off-shell. On-shell, $P^2 = m_\infty^2$, where $m_\infty\sim gT$ is
the thermal mass of a quark of momentum $\sim T$. To estimate the
virtuality of the intermediate quark, we work in the frame in which
the photon four-momentum is $K=(\omega,0,0,k_z=\omega)$; then
\begin{eqnarray}
         (P+K)^2-m_\infty^2=2P\cdot K= 2\omega(\sqrt{p_z^2+m_\perp^2}-p_z)
      \label{virtue}
\end{eqnarray}
where $m_\perp^2\equiv p_\perp^2+m_\infty^2$.  The right side of
Eq.~(\ref{virtue}) becomes very small when $m_\perp^2 \ll p_z^2$, as occurs
for small mass and emission of the photon in the forward direction,
$p_\perp\to 0$ (collinearity).  In this limit, the diagrams in
Fig.~\ref{fig:processes-2} become singular.  The quark thermal mass, which
arises from HTL resummations on the quark lines (not explicitly shown in
Fig.~\ref{fig:processes-2}), prevents these diagrams from being truly
singular, but the region of phase space where the quark and the photon are
nearly collinear leads to a contribution $\sim T^2/m_{\infty}^2\sim
1/\alpha_s$.  Combining this contribution with the explicit $\alpha_s^2$ from
the vertices, we see that these diagrams become ${\mathcal O}(\alpha_s)$.

\begin{figure}[htb]\begin{center}
\resizebox*{!}{2.5cm}{\includegraphics{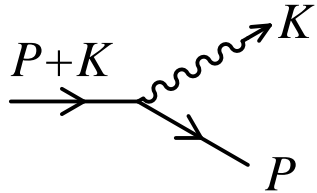}}
\end{center}
\caption{\label{fig:virtuality} A virtual quark of momentum $P+K$ emitting
a real photon ($K^2=0$) and an on-shell quark of momentum $P$
($P^2=m^2_\infty$).}
\end{figure}

    In fact, a similar collinear enhancement affects an infinite set of
processes.  The collinear enhancement in the diagrams of
Fig.~\ref{fig:processes-2}, due to the small virtuality of the quark that
emits the photon, can be rephrased physically in terms of a large {\em photon
formation time}, or equivalently, the small energy denominators in the
intermediate states.  For the process in Fig.~\ref{fig:virtuality}, the
formation time is $t_F = 1/\delta E$, with
\begin{eqnarray}
\label{dEform}
       \delta E\equiv \omega+\epsilon_{\bf p}-\epsilon_{\bf p+k} \approx
       \frac{m_\perp^2}{2}\frac{\omega}{p_z(p_z+\omega)}\; ,
\end{eqnarray}
where $\epsilon_{\bf p}=\sqrt{m_\infty^2+{\bf p}^2}$, and we assume
$m_\perp\ll p_z$.  Typically, in a quark-gluon plasma, $m_\perp \sim g T$,
while $p_z\sim T$.  Thus for a photon of energy $\omega\sim T$, we have
$\delta E\sim g^2 T$.  But $g^2T$ is in fact the characteristic scale of the
rate of collisions with small ($\sim gT$) momentum transfer $Q$.  To see this
result we write the scattering cross section as $\sigma=\int{\rm d}Q^2({\rm
d}\sigma/{\rm d}Q^2)$, where typically $ {\rm d}\sigma/{\rm d}Q^2\sim
g^4/Q^4$.  Thus the collision rate, $\gamma = n\sigma$, is $\sim
g^4\,T^3\,\int {\rm d}Q^2 /{Q^4}$, where we use $n\sim T^3$.  As we verify
later (see the discussion after Eq.~(\ref{eq:coll-simplified})), the LPM
effect regulates the integral in such a way that it becomes infrared finite
but remains dominated by the contributions of soft momentum transfers $Q\sim
gT$.  It is thus of order $1/(gT)^2$, leading to the finite result $\gamma\sim
g^4T^3/{(gT)^2}\sim g^2T$.  The formation time of a photon of energy
$\omega\sim T$ is thus of the same order of magnitude as, or larger than the
quark mean free path between two soft collisions, i.e., $t_F\sim 1/\gamma$,
the collision time.  The formation time of soft photons $\omega\sim gT$ is
even larger.

    Under such conditions, effects of multiple collisions on the production
process cannot be ignored.  Multiple scattering reduces the rate compared to
that were all collisions treated as independent sources of photon production
-- the LPM effect.  The multiple scattering diagrams that must be resummed in
the polarization tensor are the ladders in Fig.~\ref{fig:equations}.  These
processes, together with the self-energy corrections that need to be included
on the quark lines, are the typical diagrams taken into account by the
Boltzmann equation \cite{KB61} (see also \cite{Jeon95}; for a recent
derivation in the context of QCD, see \cite{Blaizot:1999xk}).

    We turn then to the explicit formulation of the photon production rate
using the linearized Boltzmann equation.  The state of the system is described
in terms of the distribution functions of charged particles, which we denote
by $n_f$ for quarks and $\bar n_f$ for antiquarks of flavor $f$ (in order to
simplify the discussion we ignore the gluons, on which the charged particle
can scatter; including their contribution poses no conceptual problem).  Our
task is then to determine the $n$'s for a system initially in equilibrium
perturbed by a weak electromagnetic potential, $A_j$; then $n_f = n_f^0+\delta
n_f$ where $n^0_f$ is the equilibrium distribution function for quarks of
flavor $f$; similarly, $\bar n_f = \bar n_f^0+\delta \bar n_f=n_f^0+\delta
\bar n_f$.  We calculate the $\delta n$'s explicitly from the linearized
Boltzmann equation.  For soft photons, $k_0,k\ll T$, this equation takes the
form,
\begin{eqnarray}
\label{kinetic1}
       (v\cdot \partial_X) \delta n_f ({\bf p},X) + e_f
       {\bf v}\cdot{\bf E} \frac{d n^0}{d\epsilon_p}= {\mathcal C}
       [\delta n_f, \delta \bar n_f];
\end{eqnarray}
here  $v^\mu=(1,{\bf v})$, with ${\bf v}={\bf p}/\epsilon_{\bf p}$, so
that $v\cdot \partial_X=\partial_t+{\bf v}\cdot {\bf \nabla}$.  In
the force term, ${\bf E}= -{\bf \nabla}A^0-\partial {\bf A}
/\partial t$ is the electric field, and $e_f$ is the charge of a
quark of flavor $f$.  (For initially isotropic distributions, the
magnetic field does not contribute to the force acting on the
particles, to lowest order.)  The $\delta \bar n_f$ are governed by
a similar equation with $e_f$ replaced by $-e_f$.  The collision
term, ${\mathcal C}$, on the right side of Eq.~(\ref{kinetic1}) is
linear in the $\delta n$'s.  The electromagnetic current $\langle
j^i\rangle$ is given in terms of the $\delta n$'s by
\begin{eqnarray}
       \langle j^i \rangle(X) = 2N_c\sum_f e_f \!\int \frac{d^3{\bf
      p}}{(2\pi)^3} v^i\left(\delta n_f({\bf p},X)-\delta\bar n_f({\bf
p},X)\right),
        \label{jn}
\end{eqnarray}
where $N_c=3$ is the number of colors, and the factor 2 accounts for the
two spin states of the quarks.

     The solution of the linearized kinetic equation is proportional to the
forcing term $\sim {\bf E}$ on the left.  It is convenient to write the
$\delta n$'s in the form,
\begin{eqnarray}
      \delta n_f({\bf p},X)&\equiv&
      -e_f W({\bf p},X) \frac{d n_f^0}{d\epsilon_{\bf
p}}=\frac{e_f}{T} W({\bf p},X)n^0_f(\epsilon_{\bf
p})(1-n^0_f(\epsilon_{\bf p})),
      \label{deltan}
\end{eqnarray}
with the same equation for $\bar\delta n_f({\bf p},X)$ with $e_f\to -e_f$.
The deviation $W$, the same for quarks and antiquarks, can be
interpreted in terms of a distortion of the local momentum
distribution caused by the shift $\delta\epsilon_{\bf p}=-e_f W({\bf
p},X)$ of the single particle energies
\cite{Blaizot:2001nr}:  $n_f({\bf p},X)=n^0_f(\epsilon_{\bf
p})+\delta n_f({\bf p}, X)=n_f^0(\epsilon_{\bf p}- e_f
W)$.  In terms of $W({\bf p},X)$, the kinetic equation reads
\beq
v\cdot
\partial_X\,W({\bf p},X) - {\bf v}\cdot{\bf E}= {\mathcal C'} [W],
\eeq
  with
  \beq\label{scaledcollision}
  {\mathcal C}\equiv -e_f(d n^0_f/d \epsilon_p){\mathcal C'}.
  \eeq

    Fourier transforming with respect to the spatial coordinates we rewrite
the kinetic equation as
\begin{eqnarray}
\label{kineticF}
       iv\cdot K\, W ({\bf p},K)+{\bf v}\cdot{\bf E}(K)=-{\mathcal C}'[W],
\end{eqnarray}
where $K$ is the four-momentum of the produced photon, and we use the same
symbol for a function and its Fourier transform, e.g., $W ({\bf p},X)$ and $W
({\bf p},K)$.  Using Eq.~(\ref{deltan}) in (\ref{jn}) we find, after
Fourier transforming,
\begin{eqnarray}\label{eq:current}
  \langle j^i\rangle(K) = -\bar e^2\int \frac{d^3{\bf p}}{(2\pi)^3}
  v^i\, W({\bf p},K) \frac{d n^0_f}{d\epsilon_p},
\end{eqnarray}
where $\bar e^2\equiv 4N_c\sum_f e_f^2$.  The extra factor 2 in
Eq.~(\ref{eq:current}), as compared to (\ref{jn}), accounts for the equal
contributions of quarks and antiquarks.  We calculate the polarization tensor
from Eqs.~(\ref{eq:current}) and (\ref{deltajA}) below.

    The linearized collision term, with Eq.~(\ref{deltan}), reads
\beq\label{gammaB}
\lefteqn{
{\mathcal C}=-\frac{e_f}{T}\sum_{f'}\int_{\bf p_1, p_2, p_3}
(2\pi)^4\delta^{(4)}(P+P_1-P_2-P_3)
\frac{\left| {\mathcal M}_{  \bf p,p_1\to p_3, p_2 }\right|^2   }
  {16 \epsilon_{\bf p}\epsilon_{\bf p_1}\epsilon_{\bf p_2}\epsilon_{\bf p_3}}
\nonumber }\\
&&\times n_f^0({\bf p})(1-n_{f}^0({\bf p_3}))n_{f'}^0({\bf
p_1})(1-n_{f'}^0({\bf p_2}))\left[W({\bf p}, K)-W({\bf
p_3},K)\right]\,.
\eeq
Here all quarks are on their mass
shells, and $\int_{\bf p_i}\equiv \int {\rm d}^3{\bf p}_i/(2\pi)^3$.  The
matrix element squared, $\left| {\mathcal M}_{{\bf p,p_1\to p_3,
p_2}}\right|^2$, is that for one-gluon exchange, as depicted in
Fig.~\ref{Born}; it is averaged over the spin and color states of the incoming
particle (of momentum $P$) and summed over the spin and color states of
the other particles.  To obtain (\ref{gammaB}) we also use the
fact that the terms involving $\delta n_{f'}({\bf p}_1,K)$ and $\delta
n_{f'}({\bf p}_2,K)$ cancel when summed over quarks and antiquarks (e.g.,
$\delta n_{f'}({\bf p}_1,K)+\delta \bar n_{f'}({\bf p}_1,K)=0$).

    At this point, it is convenient to use the four-momentum transfer ${Q}
\equiv P_2-P_1$ as an integration variable.  One can then perform the
integrations over ${\bf p_2}$ and ${\bf p_3}$, and obtain the scaled collision
term (\ref{scaledcollision}) as
\begin{eqnarray}
      \label{gammaC}\lefteqn{
        {\mathcal C}'= - \sum_{f'} \int\frac{d^4 Q}{(2\pi)^4}    2\pi
\delta(q_0-{\bf v\cdot q})  \left[W({\bf p}, K)-W({\bf p-q},K)\right]
\nonumber
}\\
&&\times\int_{{\bf p}_1}2\pi \delta(q_0-{\bf v_1\cdot
q})n_{f'}^0({\bf p_1})(1-n_{f'}^0({\bf p_1}))
   \frac{     \left| {\mathcal M}_{  \bf p,p_1\to p-q, p_1+q }\right|^2   }
{    16 \epsilon_{\bf p}\epsilon_{\bf p_1}\epsilon_{\bf p-q}
\epsilon_{\bf p_1+q}   }
\, ,
\end{eqnarray}
where ${\bf v}={\bf p}/\epsilon_{\bf p}$ and ${\bf v}_1={\bf
p}_1/\epsilon_{{\bf p}_1}$, and we have used the fact that the momentum
transfer is small ($q\ll p_i$) in order to simplify the delta functions and
the statistical factors, e.g., writing $\epsilon_{\bf p+q}-\epsilon_{\bf
p}\approx {\bf v}\cdot {\bf q}$, and $n_{f}^0({\bf p-q})\approx n_{f}^0({\bf
p})$.  After these simplifications, the integral over ${\bf p}_1$ can be done
and the contribution of the matrix element expressed in terms of the spectral
density $\rho_{\mu\nu}^{^{HTL}}(Q)$ of the gluon propagator in the HTL
approximation (see Fig.  \ref{Born}, and Ref.~\cite{Blaizot:1996az} for
details).  One then arrives at the scaled collision term
\begin{eqnarray}
      \label{gammaC2}
        {\mathcal C}'\!=- g^2 C_f\int\frac{d^4 Q}{(2\pi)^3}
   \delta(q_0-{\bf q}\cdot{\bf v})\frac{T}{q_0} v^\mu v^\nu
   \rho_{\mu\nu}^{^{HTL}}(Q) [W({\bf p},K)-W({\bf p-q},K)],
\end{eqnarray}
where $C_f=(N_c^2-1)/2N_c$.

    In order to proceed further, we examine the specific angular dependence of
the fluctuation $W({\bf p},K)$ involved in the emission of soft real photons
(typically with $\omega\sim gT \ll T$).  We first note that by
symmetry the solution of Eq.~(\ref{kineticF}) with (\ref{gammaC2}) must be of
the form
\beq
   W({\bf p},K)= {\bf v \cdot \hat E}f({\bf \hat p \cdot \hat k},p),
\label{symm}
\eeq
where $\bf \hat E$ is the unit vector along the direction of the electric
field.  This angular structure is illustrated by the collisionless Boltzmann
equation, (\ref{kineticF}) with ${\mathcal C}'=0$, which has the solution
\beq
\label{Wphotons}
    W^{(0)}({\bf p},K)=-\frac{{\bf v \cdot E}(K)}{iv\cdot K}.
\eeq
For real photons, $v\cdot K=\omega(1-{\bf v}\cdot \hat{\bf k})$, with
$\hat{\bf k}={\bf k}/\omega$.  In the case of massless particles, $v\cdot K$
vanishes when ${\bf p}$ is parallel to ${\bf k}$, leading to a diverging
$W^{(0)}({\bf p},K)$ -- the ``collinear enhancement'' discussed earlier.
Indeed, for soft photons the drift term is simply the energy
difference $\delta E$ defined in Eq.~(\ref{dEform}), i.e., $v\cdot K\approx
\omega {m_\perp^2}/{2p^2}$ (with $m_\perp^2={\bf p}_\perp^2+m_\infty^2$, ${\bf
p}_\perp \perp {\bf k}$, and ${\bf p\cdot \hat k}\gg p_\perp,m_\infty$).
Thus, noting that ${\bf E}(K)=i\omega {\bf A}(K)$, we have:
\beq
\label{collenhanced}
    W^{(0)}({\bf p},K) = - \frac{A\sin\theta\cos\phi}{2(1-\cos\theta) +
      m_\infty^2/p^2}
   \simeq - A\frac{\theta\cos\phi}{\theta^2+m_\infty^2/p^2},
\eeq
for small $\theta$,
where $\theta$ is the angle between ${\bf p}$ and ${\bf k}$, and $\phi$ the
angle between ${\bf E}$ and ${\bf p}$.  In the case of massless particles,
Eq.~(\ref{collenhanced}) exhibits the small angle divergence $W^{(0)}\sim
1/\theta$ mentioned above.  For massive particles, $W^{(0)}$ vanishes at
$\theta=0$, but remains peaked at small $\theta\sim \theta_0=m_\infty/p\sim
g$.

    Collisions, dominated by small angle scattering, maintain the peaking of
the LPM fluctuations $W({\bf p}, K)$ at small forward angles.  Accordingly the
solutions $W({\bf p},K)$ of the linearized Boltzmann equation are of the form
$W({\bf p},K)= {\bf v \cdot \hat E}h(\theta,p)$, with $h$ strongly peaked at
small $\theta$.  This structure simplifies the calculation of the collision
term, as we now show, and confirms the kinematical approximations that we made
in deriving Eq.~(\ref{gammaC2}).  Since in a collision $|\bf{p-q}|$ differs
from $|\bf p|$ by subleading terms, the magnitude of ${\bf p}$ remains
basically constant during collisions, with the direction of the dominant ${\bf
p}$ remaining approximately aligned with the momentum of the photon.  Because
$\bf q$ is primarily transverse to $\bf p$, we can neglect the dependence of
$W$ on $q_z$.  Thus $ {\bf q}\cdot \hat{\bf p}\approx q_z$, which allows us to
integrate over $q_0$ and $q_z$ in Eq.~(\ref{gammaC}).  Using the sum rule in
\cite{AurenGZ4} we find,
\begin{equation}
 \int \frac{dq_0dq_z}{2\pi} \delta(q_0-q_z) \frac{v^\mu v^\nu}{q_0}
 \rho_{\mu\nu}^{^{HTL}}(Q)
 =\frac{1}{q_\perp^2}-\frac{1}{q_\perp^2+m_{_D}^2}\; ,
\end{equation}
from which we obtain,
\begin{eqnarray}
 {\mathcal C}'=-g^2C_f T \int\frac{d^2 q_\perp}{(2\pi)^2}
 \frac{m_{_D}^2}{q_\perp^2(q_\perp^2+m_{_D}^2)} [W(\bf p)-W({\bf
  p-q_\perp})]\;.
\label{eq:coll-simplified}
\end{eqnarray}
We do not explicitly indicate the dependence of $W$ on $K$.  Since $W({\bf
p})-W({\bf p-q})$ vanishes smoothly as ${\bf q}\to 0$, the integral in
Eq.~(\ref{eq:coll-simplified}) is infrared convergent.  We expect $W({\bf
p}-{\bf q})$ to decrease rapidly when $|p_\perp-q_\perp|\gg m_\infty$, as in
the collisionless case, Eq.~(\ref{collenhanced}).  Thus the integrand in
Eq.~(\ref{eq:coll-simplified}) is dominated by momenta of order $m_D\sim
m_\infty\sim gT$, and expect the integral to remain of order $g^2T$.\footnote
{Although the electrical conductivity can be obtained as the limit of the
polarization tensor as $\omega, {\bf k}\to 0$, we cannot directly use
Eq.~(\ref{eq:coll-simplified}) to derive this  limit.  The peaking at
small angles in the LPM effect is in contrast to that in transport
calculations \cite{Baym:1990uj}, such as the electrical conductivity
\cite{Baym:1997gq,Arnold:2000dr}.  There, a similar cancellation of the small
${\bf q}^2$ contributions in the term $W({\bf p})-W({\bf p}-{\bf q})$ makes
the collision integral converge in the infrared, but the angular dependence of
the fluctuation $W({\bf p})$ induced by a uniform electric field does not
constrain the momentum transfer to be soft, and we may not neglect the $q_z$
dependence of $W$ as in deriving Eq.~(\ref{eq:coll-simplified}).  The
collision term involved in the conductivity is proportional to
$g^4T\ln(T/gT)$, where the upper cutoff in the logarithm comes from the limit
of validity of the soft momentum approximation ($q\simle T$), while the lower
one originates from screening \cite{Baym:1997gq}.  We defer discussion of the
transport solutions of the Boltzmann equation to a future publication.}

    We now show that the present Boltzmann equation, with the collision term
(\ref{eq:coll-simplified}), leads to the formulation of the LPM effect
of Arnold et al.  \cite{ArnolMY123}.  We first observe that the kinetic
equation that appears in \cite{ArnolMY123} is for a vertex function rather
than for a particle distribution.  The quantity ${\bf f^*}$ of
\cite{ArnolMY123} is related to the present $W$ by
\begin{eqnarray}
      \frac{\delta W({\bf p}_\perp)}{\delta A_j(Q)}\equiv
     - i \frac{\omega}{2p}{ f}^{j*}({\bf p}_\perp)\; ,
\end{eqnarray}
where * denotes the complex conjugate.  In terms of $f$,
\begin{eqnarray}
       {\rm Im}\,\Pi_{\rm ret}^{ij}(Q)= \frac{\bar e^2}{2\pi}
\int_{0}^{\infty}\!\! d\epsilon_p
       \,\frac{d n^0_f}{d \epsilon_p}\;
        {\rm Re}\int \frac{d^2{\bf p}_\perp}{(2\pi)^2}\;
         \omega\frac{{\bf v}^i}{2p_z} f^{*j},
\label{ImaginaryPi}
\end{eqnarray}
which coincides with Eq.~ (2.1) of the second of Refs.~\cite{ArnolMY123}.
Furthermore, the function ${\bf f^*}$ obeys the kinetic equation we obtain by
taking the functional derivative of the linearized Boltzmann equation with
respect to $A_j$.  This equation is identical to Eq.~(2.2) of the second of
Refs.~\cite{ArnolMY123}, after identification of the energy difference $\delta
E$ with the drift term in our Boltzmann equation (see the discussion below
Eq.~(\ref{Wphotons})).

    As an illustration of how the LPM effect emerges from the present
Boltzmann equation, we write, following Migdal \cite{Mig56}, a diffusion
approximation for the collision term (\ref{eq:coll-simplified}).  This
approximation is not quantitatively useful for describing quark-gluon plasmas,
since, as we shall see shortly, it requires $m_\infty\gg m_D$, a condition
which is not realized in quark-gluon plasmas where rather $m_\infty\simle m_D$
($m_\infty^2=g^2 T^2 C_f/4$, and $m_D^2=(2N_c+N_f)g^2T^2/6$).  However, it
yields simple analytical expressions which allow us to illustrate certain of
the physical points made in the preceding discussion.  The diffusion
approximation is derived by expanding $W({\bf p-q_\perp})$ to second
order in ${\bf q}_\perp$:
\bea
      {\mathcal C}'=g^2C_f T \int\frac{d^2 q_\perp}{(2\pi)^2}
      \frac{m_{_D}^2}{q_\perp^2(q_\perp^2+m_{_D}^2)}
      \frac12 ({\bf q}_\perp\cdot\nabla_{\bf p})^2 W({\bf p})
      = p^2 D \nabla_{p_\perp}^2 W({\bf p}),
\label{diff}
\eea
where the diffusion constant is $D = (g^2C_f T m_D^2/8\pi p^2 )
\ln(q_{max}/m_{_D}) $, and $q_{max}\sim m_\infty$.  That $m_\infty$ is the
appropriate upper cutoff can be understood from the following argument.  Since
$W({\bf p})$ decreases rapidly when $p_\perp\gg m_\infty$ (cf.
Eq.~(\ref{collenhanced})), we conclude that the integrand in ${\mathcal C}'$
in Eq.~(\ref{eq:coll-simplified}) is approximately constant when
$q_\perp\simle m_D$ and it behaves as $1/q_\perp^2$ for $m_{_D}\ll q_\perp\ll
m_\infty$, and as $1/q_\perp^4$ for $q_\perp\gg m_\infty$.  Thus
$q_\perp\simeq m_\infty$ is the appropriate upper bound for the integration
over $q_\perp$, and the expansion of the collision term involved in the
diffusion approximation may be viewed as an expansion in $m_{_D}/m_\infty$.
Note that the assumption $m_{_D}\ll m_\infty$ justifies the leading log
approximation used in estimating $D$.

    The Boltzmann equation in the diffusion approximation then reads
\bea
\label{kinetictheta}
    \frac{i\omega}{2}({\bf v_\perp}^2+m_\infty^2/p^2) W({\bf p})+
    D \nabla_{v_\perp}^2 W({\bf p})=-i\omega {\bf v\cdot  A}.
\eea
Again, by symmetry the solution is of the form $W({\bf p}_\perp)={\bf
v_\perp\cdot A}\,\varphi(s)/s$, where $s\equiv {\bf v}_\perp^2/2 \simeq
(\sin^2\theta)/2$.  The equation for $\varphi(s)$ is
\bea \label{kinetics}
 i\omega(s+s_0) \varphi(s)+2 s D\frac{d^2}{d s^2} \varphi(s)=-i\omega s,
\eea
with $s_0\equiv m_\infty^2/2p^2$.  In terms of $\varphi$, the current
(\ref{eq:current}) is
\beq
\label{eq:current2}
 j^i=-{\bar e}^2
 \int\frac{d^3p}{(2\pi)^3}\frac{dn^0}{d\epsilon_p}\varphi(s)
 A^i=-\frac{{\bar e}^2}{2\pi^2}\int_0^\infty p^2
 dp\frac{dn^0}{d\epsilon_p}\int_0^\infty ds \,\varphi(s) A^i,
\eeq
where we use ${\bf k\cdot j}={\bf k\cdot A}=0$, and the fact that
$\varphi(s)$ is a rapidly decreasing function of $s$, in order to extend the
range of the $s$-integration to $+\infty$.  From Eqs.~(\ref{eq:current2}) and
(\ref{deltajA}) (with $g_{ij}=-\delta_{ij}$) we derive
\begin{eqnarray}
    {\rm Im}\,\Pi_{\rm ret}^{ii}(Q)\approx \frac{{\bar e}^2}{2\pi^2}
   \int_0^\infty dp \,\frac{dn_0}{dp} \,p^2\, \Phi(\omega;p ),
\label{eq:Pi_diffusion}
\end{eqnarray}
where $\Phi(\omega;p)\equiv \int_0^\infty ds \,{\rm Im}\varphi(s)$;
the dependence of $\varphi$ on $p$ (not indicated explicitly)
comes from the diffusion constant $D\sim 1/p^2$ and $s_0=m_\infty^2/2p^2$ in
Eq.~(\ref{kinetics}).

    Solving Eq.~(\ref{kinetics}) by iteration, we derive the solution as an
expansion in powers of $D$, i.e., in the number of collisions.  In zeroth
order (no collisions), $\varphi^{(0)}=-s/(s+s_0)$, which is real and does not
contribute to $\Phi$:  there is no radiation in the absence of collisions.
Substituting the lowest order result into the collision term, we find the
single collision contribution, $\varphi^{(1)}= 4iDs s_0/(\omega(s+s_0)^4)$,
which is imaginary and yields $\Phi^{(1)}= 2D/3\omega s_0$, independent of
$p$.  Note the role of the fermion mass, entering through the factor $s_0$; as
$m_\infty\to 0$, $\Phi^{(1)}$ diverges -- the collinear divergence discussed
earlier.  Using this expression for $\Phi^{(1)}$ in
Eq.~(\ref{eq:Pi_diffusion}), we recover the low frequency Bethe-Heitler rate
from Eq.~(\ref{rate-2}):
\bea
\frac{dN_\gamma^{BH}}{d^4 x}= \frac{\bar e^2 T^3}{12\pi^2}\Phi^{(1)}{d\omega}
 =C_{BH}\frac{d\omega}{\omega},\qquad\qquad
  C_{BH}= \frac{\bar e^2 g^2 C_f T^4}{72\pi^3}
 \frac{m_{_D}^2}{m_\infty^2}\ln\frac{m_\infty}{m_{_D}}.
\eea
It is easily verified that this expression agrees with that obtained from
the formulae in Sec. 4.2 of Ref.~\cite{AurenGZ4} (in the appropriate limit,
$m_\infty\gg m_D$), an {\it a posteriori} justification for our use of
$q_{max}=m_\infty$ in Eq.~(\ref{diff}).  The emission rate is of order $g^2$
and falls as $1/m_\infty^2$.

    Proceeding further, we find that $\Phi^{(2)}=0$ and $\Phi^{(3)}\sim
D^3/(\omega^3 s_0^5)$.  The iterative solution breaks down when
$\Phi^{(3)}\simeq \Phi^{(1)}$, which occurs when $\omega\simle D/s_0^2$.  In
the small $\omega$ regime, we must solve Eq.~(\ref{kinetics}) more accurately.
The exact solution is in fact known; since Eq.~(\ref{kinetics}) is identical
to Eq.~(44) of Ref.~\cite{Mig56}, we exploit the analysis there write the
exact solution in the form $\Phi(\omega)=({2D}/{3s_0\omega})\phi(\tau)$, where
$\tau\equiv (s_0/4)\sqrt{\omega/D} $. The function $\phi(\tau)$ satisfies
$\phi(\tau\to 0)\approx 6\tau$ and $\phi(\tau\to\infty)=1$.  Thus, when
$\omega\gg D/s_0^2=D 2p^2/m_\infty^2$, $\Phi(\omega)\approx 2D/3s_0\omega$,
and one recovers the Bethe-Heitler limit.  On the other hand, as $\omega\to
0$, $\Phi(\omega) \sim \sqrt{D/\omega}$, and we see that the rate is
suppressed by a factor $\sim\sqrt \omega$, the LPM effect.  Then
\begin{equation}
  \frac{dN_\gamma^{LPM}}{d^4 x}=\sqrt{\frac{\omega}{\omega_c}} \;
  \frac{dN_\gamma^{BH}}{d^4 x} = C_{BH}\frac{d\omega}{\sqrt{\omega\omega_c}},
\label{LPM}
\end{equation}
where
\begin{equation}
\omega_c=\frac{ \pi^3 g^2 C_f T^3}{162 (\ln 2)^2}
\frac{m_{_D}^2 }{m_\infty^4} \ln\left(\frac{m_\infty}{m_{_D}}\right).
\end{equation}
Here we obtain the particular form of the spectrum $\sim
d\omega/\sqrt{\omega}$ from the diffusion approximation.  However,
the same form of the spectrum emerges, as we find, from a numerical solution
of the Boltzmann equation in the regime, $m_\infty\approx m_{_D}$, where the
diffusion approximation is no longer valid.  The diffusion approximation
primarily affects the overall normalization, and has little effect on the
shape of the spectrum for small photon energies.

    In this Letter we have considered only the case of soft photons.  More
generally -- and in particular for hard photons with $\omega\sim T$ -- neither
the energy difference $\delta E$, nor the coupling between the quarks and the
applied electromagnetic field, can be approximated in a gradient expansion.
However the kinematical conditions that allow one to obtain the linearized
collision term from the quantum field equations still hold.  The resulting
Boltzmann equation takes a similar form, but with more accurate drift and
Vlasov terms (see, e.g., \cite{KB61}).

    Authors GB and TM are grateful for the hospitality of the Aspen Center for
Physics, and GB, FG and TM to the ECT* in Trento, where part of this work was
carried out.  Authors GB and JPB thank the Japan Society for the Promotion of
Science for grants that further enabled the present research.  Grants-in-Aid
of the Japanese Ministry of Education, Culture, Sports, Science, and
Technology No.~13440067 supported TM, and No.~15740137 to Professor Tetsuo
Hatsuda of University of Tokyo supported GB.  This research was supported in
part by U.S.~NSF Grant PHY03-55014.

\end{document}